\documentclass[fleqn,10pt]{wlscirep}
\usepackage[utf8]{inputenc}
\usepackage[T1]{fontenc}

\usepackage{booktabs}
\usepackage{multirow}
\usepackage{longtable}
\usepackage{xr-hyper}
\usepackage[charter,cal]{mathdesign}

\makeatletter
\newcommand*{\addFileDependency}[1]{
  \typeout{(#1)}
  \@addtofilelist{#1}
  \IfFileExists{#1}{}{\typeout{No file #1.}}
}
\makeatother

\newcommand*{\myexternaldocument}[1]{%
    \externaldocument{#1}%
    \addFileDependency{#1.tex}%
    \addFileDependency{#1.aux}%
}
\myexternaldocument{si}

\title{
American cities are defined by isolated rings and pockets characterized by limited socio-economic mixing
}

\author[a,*]{Andrew Renninger}
\author[a]{Neave O'Clery}
\author[a]{Elsa Arcaute}

\affil[a]{Centre for Advanced Spatial Analysis, University College London, London, UK}

\affil[*]{Corresponding author: Andrew Renninger (E-mail: andrew.renninger.12@ucl.ac.uk)}

\begin{abstract}
Cities generate gains from interaction, but citizens often experience segregation as they move around the urban environment. Using GPS location data, we identify four distinct patterns of experienced segregation across US cities. Most common are affluent or poor neighborhoods where visitors lack diversity and residents have limited exposure to diversity elsewhere. Less frequent are majority-minority areas where residents must travel for diverse encounters, and wealthy urban zones with diverse visitors but where locals sort into homogeneous amenities. By clustering areas with similar mobility signatures, we uncover rings around cities and internal pockets where intergroup interaction is limited. Using a decision tree, we show that demography and location interact to create these zones. Our findings, persistent across time and prevalent across US cities, highlight the importance of considering both who is mixing and where in urban environments. Understanding the mesoscopic patterns that define experienced segregation in America illuminates neighborhood advantage and disadvantage, enabling interventions to foster economic opportunity and urban dynamism.
\end{abstract}
\begin{document}

\flushbottom
\maketitle

\section*{Introduction}
The free flow of people, goods and ideas is a key component of any urban agglomeration. Cities enable "sharing, learning and matching" to foster efficient and innovative economies \cite{duranton2004micro}: shared infrastructure, shared ideas, and pooled workers drive the dominance of urban systems in the economy today. By extension, cities that fail to integrate communities or struggle to connect residents may sacrifice growth \cite{harari2020cities} or exacerbate poverty \cite{ananat2011wrong, cutler1997ghettos}. A large body of literature documents the ways in which residential communities within cities segregate by race and class, and recent developments, including the availability of location data derived from mobile phone pings, allow us to investigate segregation in daily life as people move around the city \cite{athey2021estimating, moro2021mobility}. Better understanding of the ways in which different groups interact is key to unlocking innovation and prosperity in cities \cite{ribeiro2023mathematical, bettencourt2013origins, bettencourt2007growth}.

Cities in the United States exhibit marked patterns of ``experienced segregation", the propensity for urban residents to sort according to socio-economic attributes during their day-to-day activities. People are more likely to share spaces with others of the same race \cite{athey2021estimating} and class \cite{moro2021mobility}. Recent work emphasizes that venues that are at short distances apart can have distinct visitor profiles \cite{moro2021mobility}, with both spatial factors, like transit access, and social factors like the demographic composition of a venue's clientele, influencing who visits it \cite{davis2019segregated}. Studies also indicate that the location of amenities influences daily mobility patterns \cite{heroy2023neighbourhood} and may also mediate interactions between groups \cite{nilforoshan2023human}: if a city has clusters of amenities between rich and poor communities, those communities will interact more. There is also evidence that patrons prefer to visit businesses frequented by those in the same economic stratum, but that when mixing between strata occurs, it is usually with people from a lower class visiting a business popular with those from a higher class \cite{hilman2022socioeconomic, noulas2012tale}. 

While these studies measure exposure rather than true social interaction \cite{zhou2019between}, a variety of studies link mobility with social ties and their economic consequences. Activity spaces—the bounds of our routines in space—influence who our friends and acquaintances are \cite{kovacs2023income, bokanyi2021universal}, with particular consequences for ``weak ties" \cite{granovetter1973strength} which help generate social and economic mobility. Employment opportunities within an individual's activity space predicts whether or not they become employed \cite{cagney2020urban}. Traditional measures of neighborhood segregation are associated with social capital and with it social mobility \cite{chetty2022social1, chetty2022social2}, but recent work finds that areas with venues that encourage mixing also tend to have greater social capital and social mobility \cite{massenkoff2023rubbing}. All of this indicates that generating connections between groups and across communities may create wealth and reduce inequality. 

Although the literature that describes experienced segregation is growing \cite{liao2024socio}, studies tend to focus on human behaviors \cite{moro2021mobility, yabe2022behavioral, bokanyi2021universal, hilman2022socioeconomic, massenkoff2023rubbing, cook2022urban} or urban comparisons \cite{athey2021estimating, toth2021inequality, abbiasov202215, nilforoshan2023human}. We can think of these as microscopic and macroscopic views of experienced segregation, with few considering the mesoscopic zones within and around cities that drive mixing or isolation and the forces that generate these zones. Early work on concentrated poverty contends that limited connections to the broader economy create ``truly disadvantaged" communities, i.e., those with multiple coincident forms of deprivation \cite{wilson2012truly}, and later work shows that new connections might ameliorate some of these problems \cite{chetty2016effects, chetty2018impacts}. The location and concentration of experienced segregation reveals not just who is isolated on an individual level but also who is surrounded by isolated groups, and thus less likely to develop these new connections. Hence, discerning the clusters of experienced segregation will allow us to identify the most disadvantaged communities, and propose targeted interventions.

The following work looks at the broader zones of relative mixing and isolation that define cities. Zones where residents either avoid contact with other groups or seek it out convey important information about inequalities between groups, as certain groups or places may make uneven contributions to the aggregate socio-economic mixing in a city. Our contribution is threefold. Building on these macroscopic and microscopic perspectives, this work investigates the mesoscale of mixing in America. By exploring a spatially and temporally expansive sample of cell phone GPS records covering the continental United States over 4 years, we explore the prevalence and durability of these patterns. Finally, we propose a new way to understand contributions to intergroup contact that classifies the role of a neighborhood within its city, distinguishing between communities that contribute to mixing and those that avoid it. 

We present measures of place \emph{segregation} and neighborhood \emph{isolation} for the United States, representing the degree to which visitors to a place are diverse (\emph{segregation}), and the degree to which travelers from a neighborhood experience diversity over the course of the day-to-day activities (\emph{isolation}). Considering these variables together, we see that they are related but distinct, and the combination of them produces classes that define neighborhoods within the broader urban system. We document cases where segregation and isolation are at variance, including neighborhoods that export diversity to the broader city without importing it, and use clustering to understand the larger zones of segregation and isolation that define American cities.  

\begin{figure}[t!]
\centering
\makebox[1\textwidth][c]{\includegraphics[width=1\columnwidth]{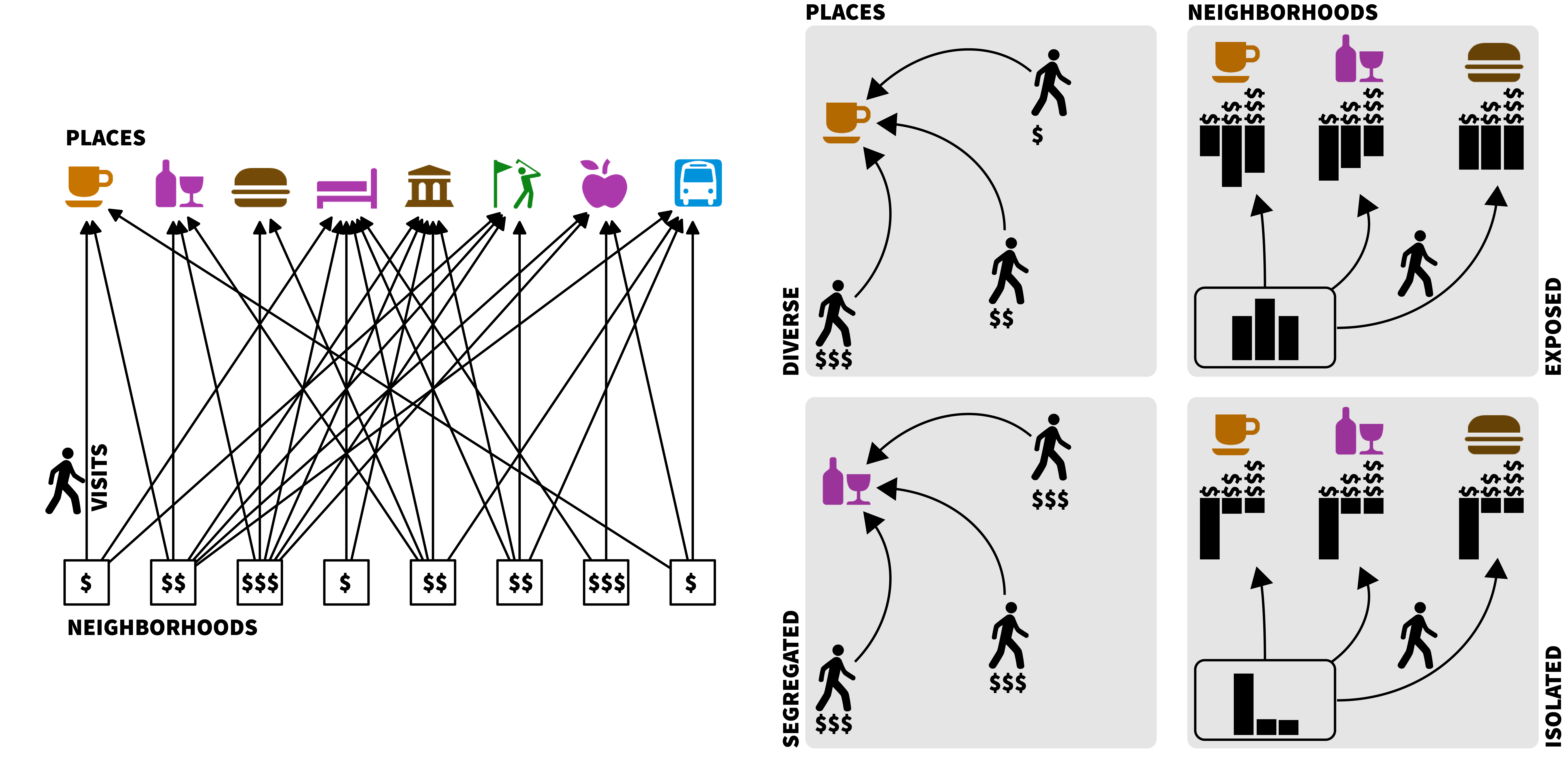}}
\caption{\textbf{Illustration of how we measure segregation and isolation}. We begin with visits to places, with visitor income imputed from Census data, and then use the diversity profiles of those places to compute the exposure of neighborhoods that sent visitors to them.} 
\label{diagram}
\end{figure}

\section*{Results}

\subsection*{Data and methods}
To assess experienced segregation across multiple dimensions, we construct measures for diversity and exposure in terms of socio-economic class and race. We process data from a location intelligence provider \cite{safegraph2019visitation}  to create origin-destination flows from neighborhoods (Census block groups) to points of interest (restaurants, shops and other amenities) over a period from January 2019 through December 2022. We then combine these data with Census estimates of median income \cite{USCensusBureau_ACS} to infer the socio-economic strata of visitors from each area by attributing to each visitor the median income of the block group where they live. We compute the ``diversity" of visitors to points of interest and the ``exposure" of neighborhood residents to diversity. Borrowing from prior work \cite{moro2021mobility}, our measures represent distance from a counterfactual scenario where visitors to a place draw equally from each socio-economic stratum in a given city. We examine place segregation (diversity of visitors to a place) and neighborhood isolation (exposure of residents to diversity in the places they visit), considering a place highly segregated if it attracts visitors from just one income bracket, and residents highly isolated if they only visit such places. Both measures range from [0, 1], where 0 is perfect diversity or exposure and 1 is perfect segregation or isolation. We illustrate our process in Fig. \ref{diagram}; see \textbf{Methods} for more detail.  

\subsection*{Segregation and isolation}

\begin{figure*}[bt!]
\centering
\includegraphics[width=1\textwidth]{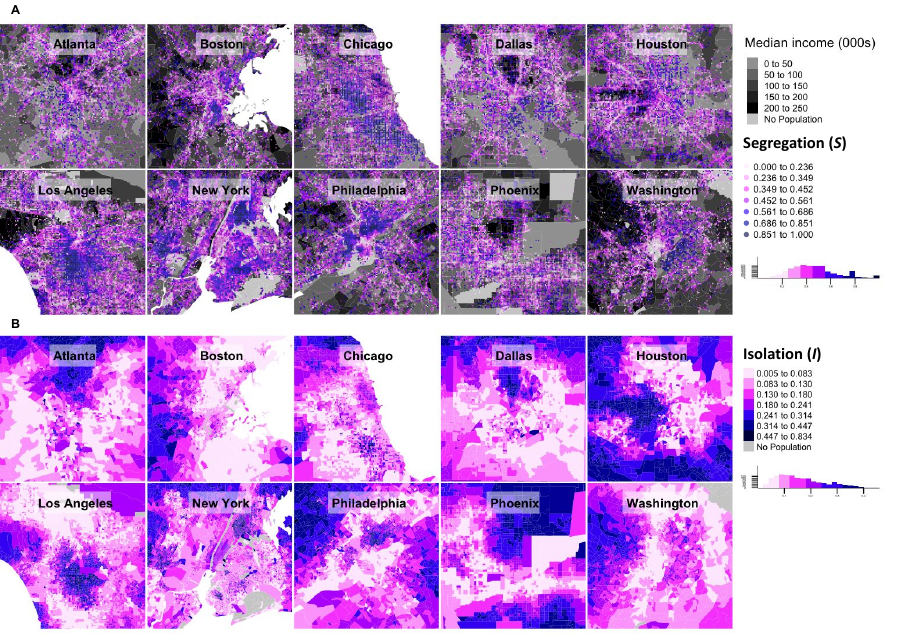}
\caption{\textbf{Measures of place segregation (\emph{S}) and neighborhood isolation (\emph{I}).} \textbf{A} Place segregation, where each point represents a POI and each city is centered on downtown, shows that downtown businesses tend to see a diverse collection of visitors (and thus have low place segregation) but that businesses in surrounding neighborhoods often do not (and have high place segregation). Many of the wealthiest parts of the city, shown in black, also have fewer points of interest, which limits visitation and thus the diversity. \textbf{B} Neighborhood isolation is strong in those same wealthy areas with fewer POIs and also in areas with segregated POIs.}
\label{univariatemaps}
\end{figure*}

We can think of a neighborhood as isolated and a place, or point of interest, as segregated; while isolated neighborhoods may have segregated POIs, a neighborhood and its POIs are separate and can in principle have distinct mixing profiles. Fig. \ref{univariatemaps} displays these measures of segregation and isolation for 10 large cities. While previous work \cite{moro2021mobility} has emphasized variability between nearby amenities of interest and within neighborhoods, showing that a pair of adjacent POIs may have opposite diversity signatures, Fig.~\ref{univariatemaps} shows that these form distinct clusters of both segregation and diversity. Our sample includes cities with different spatial structures, from sprawling ``sunbelt" cities like Houston and Dallas to dense ``rustbelt" cities like Chicago and Philadelphia. The latter tend to have pockets of segregation, like North and West Philadelphia, the south side of Chicago, and the South Bronx in New York. In contrast, Houston and Dallas are more integrated. Los Angeles runs counter to the distinction between colder, older metros with concentrations of limited mixing and warmer, newer metros with blurrier demarcations: it has a pocket of segregation in South Los Angeles—a historically disadvantaged neighborhood. 

One channel by which segregation is determined is in the mix and number of amenities in a cluster: fewer reasons to visit will limit visitors. It is clear from the locations and concentrations of points of interest that many wealthy areas—highlighted in black—limit visitation simply by limiting the quantity of places to visit. We see examples of this in Dallas, Phoenix, Washington, and Los Angeles—corresponding with Bel-Air and other wealthy enclaves.   

To understand the relationship between segregation and isolation, we bin each by quantile and construct a 3x3 matrix that classifies each neighborhood by its performance on both measures. That is, a neighborhood that is completely integrated, with diverse amenities and with residents visiting diverse places, will be in the first quantile along both dimensions—diversity and exposure, in and out respectively. Vice versa, homogeneous amenities and residents visiting homogeneous places will be in the third quantile for both. Residents may also find exposure to diversity away from home, and neighborhoods with them will be in a low quantile for isolation and a  high quantile for segregation. Further, because we capture segregation at the level of the point of interest, it is not a foregone conclusion that an area with diverse amenities will have residents exposed to diversity: they can seek out amenities with a given demographic profile even as the neighborhood around them receives a different selection of visitors.

\begin{figure*}[bt!]
\centering
\includegraphics[width=1\textwidth]{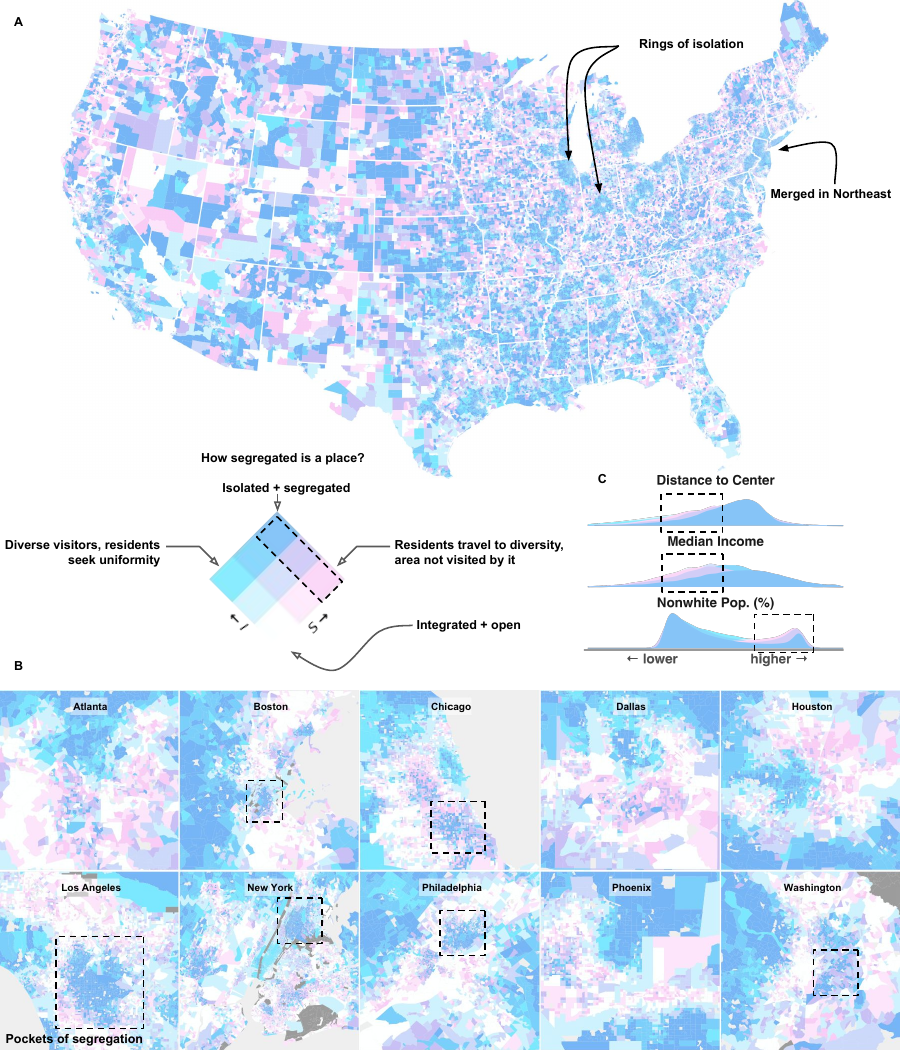}
\caption{\textbf{The relationship between segregation and isolation.} \textbf{A} Diversity and exposure nationally we see that integrated urban areas are often surrounded by rings of isolated suburban areas. \textbf{B} Locally, we see urban pockets of segregation with a range of low to high isolation, often near or surrounded by integrated urban areas. \textbf{C} Distributions of select variables by class show that segregated areas are often close the center, poorer than average, and nonwhite.}
\label{bivariatemaps}
\end{figure*}

We combine diversity and exposure in Figs. \ref{bivariatemaps}A and \ref{bivariatemaps}B, nationally and locally. The resulting map shows that many cities are surrounded by rings with high segregation and high isolation. In the Boston-Washington megalopolis, many of these rings merge into large intercity zones. Looking within cities, we see that there are areas where visitors are diverse but exposure is low: residents sort into points of interest that allow them to avoid ambient diversity. These neighborhoods as heterophobic, and they are typically central and affluent. The opposite also occurs, often in the same cities: heterophilic neighborhoods in Chicago and New York have residents that experience diversity despite living in a neighborhood that sees little of it. We also see that historically deprived neighborhoods are often both segregated and isolated. To understand these neighborhoods, we plot distributions of select variables by class in Fig.~\ref{bivariatemaps}C: these neighborhoods which export diversity without importing it tend to exist at a certain distance from downtown, with characteristic socio-economic attributes—high nonwhite populations and low incomes. (A detailed and interactive map of these data is available \href{https://asrenninger.github.io/diversity/}{here}.)        

Estimates of sergegation $S$ and isolation $I$ hold when we bin neighborhoods according to race instead of income, with generally higher values for race than for income in larger cities (see Supplementary Fig. S2); they are also systematically higher than values generated with a null model that uses commuting data rather than POI visitation to estimate $S$ and $I$ (see Supplementary Fig. S3). 

\section*{Zones of segregation and isolation}
\begin{figure*}[bt!]
\centering
\includegraphics[width=1\textwidth]{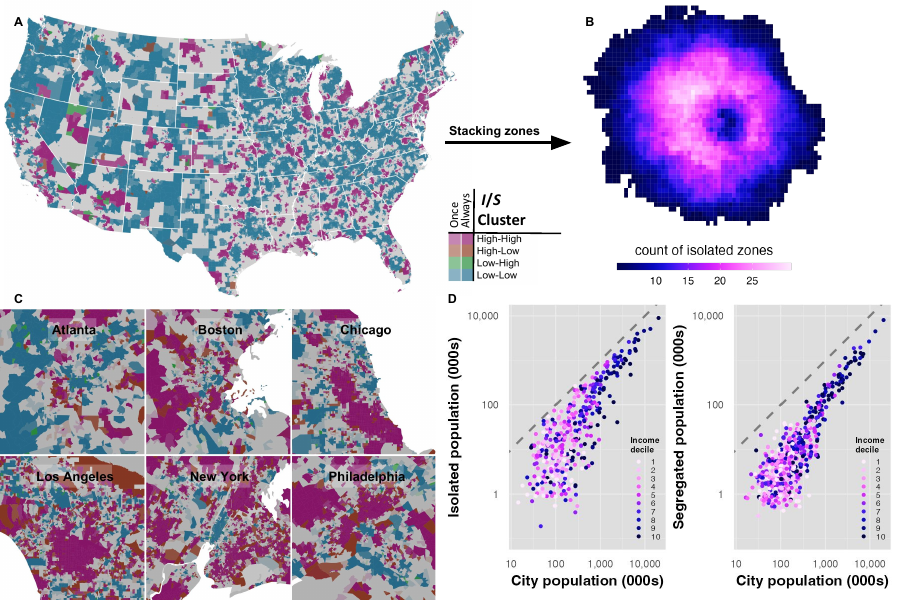}
\caption{\textbf{Defining rings of isolation and pockets of segregation.} \textbf{A} Isolation autocorrelation manifests at the national scale, delineating rings around cities. Centering and layering the cities, in \textbf{B} we count the number of times isolated zones occur in the same relative area: there is a clear prevalence of these zones in a ring surrounding each urban core. \textbf{C} Segregation autocorrelation manifests locally, with pockets appearing in large cities—less so in smaller cities like Atlanta or Boston. We also see tight scaling relationships between city size and isolation/segregated population in \textbf{D}, with a linear relationship between city population and the population in isolated zones along with a superlinear relationship for city population and its segregated zones.}
\label{lisa}
\end{figure*}

In order to understand how these measures of interaction manifest in the structure of cities, we use tests for spatial autocorrelation to identify contiguous zones of place segregation or neighborhood isolation. These tests show that at both local and regional scales there are large areas where high values cluster. Segregation and isolation exist at characteristic scales, with the latter clustering in suburban rings and the former clustering in certain urban pockets, shown in Fig. \ref{lisa}A and \ref{lisa}C, respectively. Fig. \ref{lisa}A, in particular, shows that many cities are surrounded by bands of isolated neighborhoods. In Fig. \ref{lisa}B, we center the top 100 cities so that downtown is (0,0) and then layer these rings, in effect create a synthetic metro that aggregates zones from all observed ones; we divide this synthetic metro in a grid to create synthetic neighborhoods. This allows us to ask, amongst neighborhoods that generally have this location relative to the center, how many are isolated? Counting the overlap per grid cell answers this question and shows that many rings in different cities exist at a similar location relative to downtown. Compared to clusters of isolation, which tend to be suburban, segregation tends to be urban and the clusters tend to exist at a smaller, more fragmented scale. The panels in Fig. \ref{lisa}C show that large cities—New York, Los Angeles, or Chicago—also tend to the have larger concentrations of segregation, compared to small cities like Atlanta and Boston. 

The populations of these zones scale predictably with city size, shown in Fig. \ref{lisa}D. For isolated clusters, the relationship is linear beyond a city size of 500,000: in a city of 1 million people, we expect 500,000 people in these zones; for a city of 10 million, we expect 5 million. The population in segregated clusters scales superlinearly to the population of the city. This means that as cities become larger, separated areas with limited visitation become larger at a faster rate—with large areas visible in big cities (New York, Chicago) and less so in small cities (Atlanta, Boston) in Fig. \ref{lisa}C. 

\subsection*{Trends and drivers}
\begin{figure*}[bt!]
\centering
\includegraphics[width=1\textwidth]{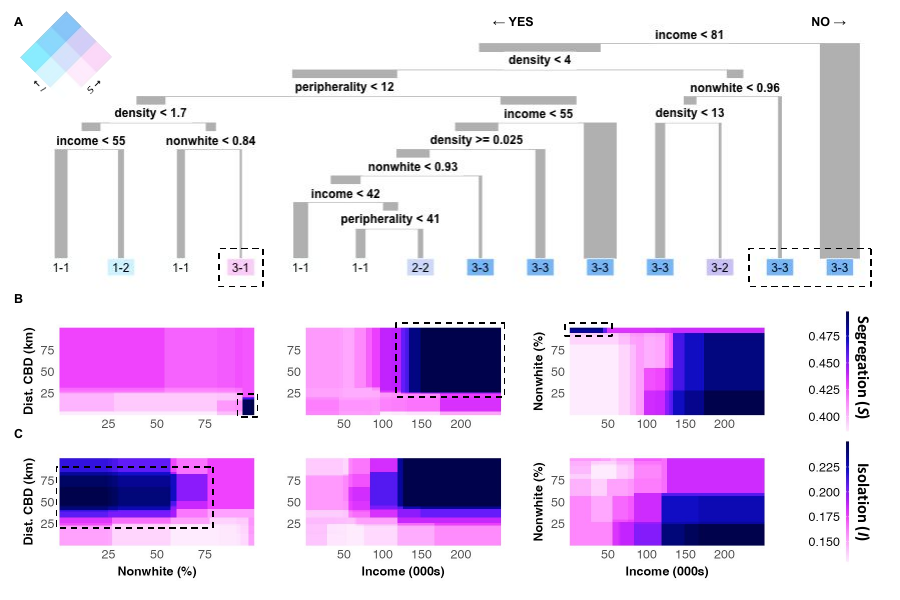}
\caption{\textbf{Factors associated with segregation and isolation.} \textbf{A} Decision tree showing the defining characteristics of different classes, pruned for ease of  viewing. Note on the right side that top segregated and isolated classes are the wealthiest cut, but also urban nonwhite—as indicated by density. The areas that have high segregation but low isolation tend to be urban, nonwhite and moderately dense. Partial dependence plots for \textbf{B} segregation and \textbf{C} isolation showing the joint relationship between key variables, with rings of isolation conditional on white/wealthy and pockets of segregation conditional on nonwhite/poor.}
\label{bivariatetree}
\end{figure*}

We turn to a decision tree, shown in Fig. \ref{bivariatetree}A, to understand what defines each class of our 3x3 scheme. We append attributes to each neighborhood using data from the Census, along with amenity data from SafeGraph. Also measure distance to the central business district (CBD) taking each block group's distance to its parent city's largest cluster of amenities, determined using DBSCAN \cite{hahsler2019dbscan}. These become predictors in our model. By combining variables to generate predictions, a decision tree helps understand nonlinear relationships; because extremes along some dimensions exhibit similar characteristics along others, a linear model is insufficient. In addition, this provides us with an intuitive, step-by-step way to see which variables predict which class, in order of importance. Points of decision include richer or poorer, densely or sparsely populated, predominately white or nonwhite, centrally or peripherally located. For example, we can ask what is the most common class of a neighborhood that is near the center, predominately nonwhite, relatively dense and relatively poor. While we could check all permutations manually, a decision tree prunes this space and keeps only what is predictive, making the process faster and simpler. 

Our complete tree, shown in the supplementary materials, achieves an accuracy of 30\%, much better than the 11\% chance when guessing 9 categories. Table \ref{importance} shows the variable importance of each feature in the tree, using a permutation technique that shuffles each column and observes the resulting reduction in predictive ability. Income, along with other measures of urban structure like density and distance from center, are the most important predictors. This ranking suggests that, along with income, urban form and function, including the distribution of amenities, may be important determinants of segregation and isolation—though we cannot rule out the role of selection, wherein people in isolated groups sort into certain neighborhoods. We show a surrogate tree—pruned to be interpretable—in Fig. \ref{bivariatetree}A. There is an interesting symmetry in the fact that cophenetically proximate branches of the tree indicate that both the wealthiest neighborhoods and dense, nonwhite neighborhoods are the most segregated. We also see that the most likely populations to have high segregation and low isolation are nonwhite. In less affluent and less white communities, the presence of amenities is associated with a greater diversity of visitors. (See Supplementary Fig. S1 for greater detail.) 

Another important aspect of the tree is the irrelevance of unemployment: insofar as a job represents a connection to the broader urban economy, we might expect working population to predict isolation above and beyond income. It does not: the best model uses income and race, along with spatial structure, to estimate isolation and segregation. Joblessness will covary with income, but it provides little additional information. This fits with the weak correlation between commutes and mixing (See Supplementary Fig. S2): where you work does not do much to help exposure if you sort into venues with an audience similar to you. 

\begin{table}[h!]
\begin{center}
    \begin{tabular}{l|r|r}
        \textbf{Rank} & \textbf{Variable} & \textbf{Importance} \\ \midrule
        1 & Median income & 2054 \\
        2 & Dist. to CBD & 1224 \\
        3 & Density & 1146 \\
        4 & Amenity (\#) & 937 \\
        5 & Nonwhite (\%) & 786 \\
        6 & College (\%) & 634 \\
        7 & Amenity ($H$) & 396 \\
        8 & Household size & 253 \\
        9 & Vacancy rate & 192 \\
        10 & Rent burden & 155 \\
        11 & Under 16 (\%) & 100 \\
        12 & Unemployed (\%) & 56 \\ \midrule
    \end{tabular}
    \caption{\textbf{Table of important features in our tree.} Income is the single best predictor of segregation but urban structural factors like distance to the CBD, density and the number of amenities are also important.}
    \label{importance}
\end{center}
\end{table}

The decision tree allows us to explore the relationship between key variables conditional on all other predictors \cite{goldstein2015peeking}, which helps to model the interactions in our data. Fig. \ref{bivariatetree}B and C show the model's estimated values for segregation and isolation given certain conditions; it asks, given these values for demographic, economic or geographic features, what do we expect the segregation and isolation to be? Here we are disaggregating our estimates of isolation and segregation, rather than looking at the combined classes above, so the information is distinct from Table \ref{importance}.   

The patterns are clear: if a place is close to its downtown, it is more integrated, unless it is majority nonwhite; if a neighborhood is far from its downtown, it is more isolated, unless it is majority nonwhite. The relationship between location within the urban system and demographic composition creates quadrants that correspond to higher or lower segregation. Our decision tree indicates that race is the most important driver of place segregation: the highest modeled estimates for place segregation come in neighborhoods with large nonwhite populations—regardless of how central the location is. Wealthy neighborhoods also tend to have more segregated places, but only if they are majority white. The rings that we see around each city are also evident in these distributions, with segregated suburbs becoming comparably less segregated exurbs at the farthest extents of the city. These rings stretch far from the urban core—up to 75km—which suggests that it is satellite cities that provide venues for mixing, rather than suburbs to major metros. That is, at those long distances, new centers emerge to support activities and these areas are more integrated. We test this by incorporating distance to subcenters, clusters of amenities with fewer points of interest than the urban center but more than 50. Close proximity to these subcenters is associated with lower estimates for segregation and isolation in our model. 

\begin{figure*}[bt!]
\centering
\includegraphics[width=1\textwidth]{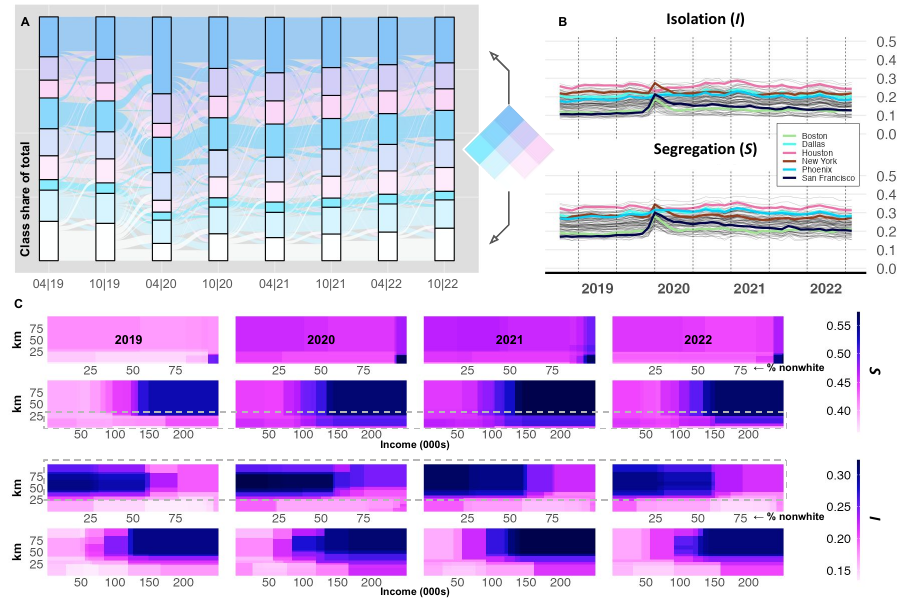}
\caption{\textbf{Factors associated with segregation and isolation.} \textbf{A} Decision tree showing the defining characteristics of different classes, pruned for ease of  viewing. Note on the right side that top segregated and isolated classes are the wealthiest cut, but also urban nonwhite—as indicated by density. The areas that have high segregation but low isolation tend to be urban, nonwhite and moderately dense. Partial dependence plots for \textbf{B} segregation and \textbf{C} isolation showing the joint relationship between key variables, with rings of isolation conditional on white/wealthy and pockets of segregation conditional on nonwhite/poor.}
\label{trends}
\end{figure*}

Looking across time, the bivariate class comprised of the most segregated and isolated areas, flowing across the top of Fig. \ref{trends}A, became larger during the pandemic but has since returned to its 2019 fraction of the total; the class comprised of least segregated, flowing across the bottom, has still not recovered—an indication that downtowns are less places of mixing than before. Confirming previous work on the subject \cite{yabe2022behavioral}, we show that experienced segregation peaked during the pandemic; yet we extend this work to show that by the end of 2022, levels of diversity and exposure had returned to 2019 levels across much of the sample. Fig. \ref{trends}B shows that, aside from San Francisco and Boston, the distribution has largely returned to its index value. This indicates that isolation and segregation have stabilized after the initial shock of the pandemic and do not appear to drifting one direction or another due to remote working patterns. 

The clusters of isolation and segregation are also stable over time, which we assess by determining the probability of an area moving from one kind of cluster to another between years in the sample. For isolation, areas that were isolated and surrounded by isolation before the pandemic have remained so: 49\% of neighborhoods both are in an isolated cluster at baseline and endline, and just 10\% have transitioned in one direction or another. Furthermore, $\sim85\%$ of all neighborhoods remained either in an integrated or an isolated cluster. Segregation is more variable: only a quarter of all neighborhoods were in a segregated cluster in both periods examined here, compared to 33\% for isolated clusters (see: Supplementary Fig. S4 for greater detail). There drift toward greater segregation, though limited, is concentrated in the Northeast and Midwest, which we map in Suppementary Fig. S5. 

In order to understand how this change and churn manifests in the structure of cities and the zones we identify above, we track how the partial dependencies from above have shifted over time in Fig. \ref{trends}C. Most predictions, either for segregation or isolation, shift to higher values throughout the study period. A notable change occurs in the band representing wealthy neighborhoods close to the urban core, which become more segregated in 2021 and 2022 despite proximity to downtown amenities that facilitated mixing in 2019. Urban areas with 90\% nonwhite populations had high levels of place segregation prior to the pandemic, as of 2022 the threshold of nonwhite residents at which our model predicts high levels of segregation is 80\%; in 2019 our model expects urban areas within 5-10km from the center with median income of \$200,000 to have $S=0.4$ and in 2022 our model expects those same areas to have $S=0.5$, a 20\% increase. The rings that we see around cities above also extend out: predictions for isolation become higher beyond 75km, where the highest predictions ended in our 2019 baseline. In these exurban areas or satellite cities greater than 75km from the city center, the average estimated isolation rises from $I=0.25$ to $I=0.35$ where neighborhoods are majority white.

 We also see little change in the rank importance of features over the period analyzed here (see Supplementary Fig. S6). This indicates that the tree is stable over time. Although it maintains its rank importance, the level of importance of distance to the CBD does drop in 2021 and 2022, indicating that we may be seeing changes to the spatial structure of segregation and isolation. The evidence presented here generally suggests that the relationships between predictors and segregation or isolation are stable in direction but often growing in magnitude.

\section*{Discussion}
We have shown here that capturing the diversity of visitors to an area and measuring the exposure to diversity of residents from an area convey distinct but related aspects of urban life. In doing so, we have identified neighborhoods at different extremes, participating in the urban economy in different ways. Suburban neighborhoods tend to be isolated, downtowns tend to be integrate. Residents from urban neighborhoods with large nonwhite populations often contribute diversity to other neighborhoods without receiving visitors from them; on the contrary, in many urban neighborhoods with wealthy and white populations—including several in large cities like New York and Chicago—visitors are diverse but residents sort into homogeneous venues, avoiding ambient diversity.

Our analysis reveals that cities are comprised of spatially contiguous zones of isolation and segregation, wherein residents interact primarily with other residents of a similar socio-economic class. These zones can often stretch for great distances, encompassing large fractions of the total population, usually in the suburban communities that surround cities. The approach we have taken to show these relationships—decision trees recalibrated over time—allows us to see that the presence of contiguous zones with limited mixing are the product of social attributes, like race and income, and urban spatial structure: the population of a place moderates any impact of its relative centrality.   

It is important to note that while we see characteristic rings of isolation around cities and pockets of segregation near the center, implicating urban spatial structure in the phenomenon, distance to the urban core is a weak predictor of neighborhood isolation or place segregation. Cities are too variable in important dimensions, like size, sprawl, or transit access, and historic contingencies may influence a neighborhood's integration into the city as a whole. Instead, the relationship between mobility and location is moderated by race and class. The combination of location within the urban system and socio-economic information accounts for most of the variation in the data. Most high values of place segregation occur when an area is both central and majority nonwhite; most high values of neighborhood isolation occur when an area is peripheral and majority white. Central and white, in affluent downtowns, typically indicates low segregation; peripheral and nonwhite is mixed.

This work builds on previous work documenting spatial patterns and temporal trends in experienced segregation. We show that there are marked spatial patterns, whether we focus on amenities or neighborhoods: while previous work has explored the limits of autocorrelation between points of interest, showing that restaurants down the street from each other can have distinct segregation profile, we see that each city has a structure that minimizes mixing in some areas while maximizing it in others. This tends to differ by region, with Southern cities like Houston and Dallas, each dependent on the automobile, exhibiting contrasting patterns to Norther cities like New York, Boston, Philadelphia, and Washington. Generally, however, we find clusters of segregation and isolation in every city—clusters that scale predictably with city size and are often stable over time. Notably, zones of isolation scale at a linear rate with respect to city population size: larger cities have commensurately larger isolated populations. By contrast, the zones of segregation tend to be smaller and are often situated near the urban core—and these scale superlinearly with city size.

With COVID-19 and remote work, networks of connections and interactions in cities are rewiring, and many of the features that defined mobility before the pandemic may not hold as urban systems equilibrate \cite{ramani2021donut, barrero2023evolution}. Research has documented changes in mobility generally \cite{santana2022changes} and experienced segregation specifically \cite{yabe2022behavioral} before and after the pandemic, finding that COVID-19 brought considerable changes to daily routines and to socio-economic mixing. The office \cite{yabe2023dependency} and the commute \cite{miyauchi2021economics} are key determinants of ``third places" \cite{oldenburg2013problem}—shops, restaurants and cafes—we visit throughout the day. Their evolving role in our daily routines will determine much of how we travel and intermix going forward. 

Our results indicate that many of the patterns dictating who interacts with whom in American cities have evolved. Trends in experienced isolation have stabilized since the shock of the pandemic, and our models are generally constant since over the past 2 years. San Francisco has seen strong changes to city life throughout the pandemic, and its decline shows in our data; other cities have returned to prior states. The determinants of segregation and isolation are also changing over time: many downtowns—especially wealthy ones—are no longer venues for mixing, suggesting that remote work may impact it durably, and many suburbs are becoming more isolated. These patterns appear strongest in the ``rustbelt"—America's industrial heartland—and weakest in the ``sunbelt" growing cities, which warrants further investigation.     

\section*{Methods}

With the goal of assessing experienced segregation along multiple dimensions, we construct measures for diversity and exposure for both socio-economic class and race. To achieve this, we use data from SafeGraph \cite{safegraph2019visitation}, a location services provider, to construct origin-destination flows from neighborhoods (Census block groups) to points of interest. SafeGraph gathers GPS locations from mobile phones by aggregating data from applications who have obtained user consent to passively monitor location, creating a sample of users comprising $\sim10\%$ of the population. Their process assigns visits to places by clustering GPS pings and joining these clusters to adjacent building polygons, using relative distances and time-of-day to manage conflicts \cite{safegraph2019visitation}. Previous work has shown that SafeGraph data are demographically balanced \cite{squire2019bias}, and the data have been used in a variety of academic contexts \cite{chen2018effect, chang2021mobility}. The result is a rectangular matrix consisting of 220,000 origin home block groups (a Census aggregation with a population $\sim1000$) and 7 million destination points of interest like restaurants, museums, cafes, car dealers, or grocery stores. To each origin we append Census estimates from the American Community Survey \cite{USCensusBureau_ACS} of median income and nonwhite population in order to infer socio-economic strata and demography of the visitors from that area. With these we can compute the ``diversity" of (visitors to) points of interest and the ``exposure" of (residents in) neighborhoods to diversity.  

\paragraph{Measuring intergroup interaction.} We look at place segregation and neighborhood isolation. The first asks, how diverse are visitors to this place? The second asks, how much diversity are residents from this neighborhood exposed to in daily routine? Segregation captures the patrons at a given place, or point of interest (POI), which we can then use to compute isolation at the level of the neighborhood by taking the weighted average of segregation at points of interest visited by the residents of a given neighborhood. The measures we construct represent deviations from a counterfactual scenario where visitors to a place draw equally from each socio-economic stratum in a given city. If a restaurant attracts visitors from just one income bracket, calibrated within that metropolitan area, we consider that place to be highly segregated; if residents from a neighborhood only go to restaurants like that one, we consider them to be highly isolated. Diversity and exposure are the inverse of segregation and isolation. We illustrate this process in Fig. \ref{diagram}.    

Following earlier work \cite{moro2021mobility}, we consider the segregation $S$ of an amenity $\alpha$ to be a distance from an ideal scenario where people from all socio-economic classes visit in equal proportions. This is defined as follows

\begin{equation} \label{eq:1}
S_{\alpha} = \frac{5}{8} \sum_{q} \left| \nu_{q\alpha} - \frac{1}{5} \right|,
\end{equation}

\noindent
where $q$ represents an income quintile and $\nu$ represents the portion of visitors from that quintile. We scale that by $\frac{5}{8}$ so that each value spans 0 to 1, with 0 being perfect integration (equal proportions from all classes) and 1 being perfect segregation (visitors from a single class). Each quintile is calibrated to the metropolitan area, rather than the nation as a whole. As a robustness check, we also compute quintiles based on the nonwhite population, differentiating between neighborhoods based on the proportion of nonwhite residents. 


Neighbourhood isolation $I$ is obtained by aggregating $S$ as follows. We compute the average diversity $S$ for all places $\alpha$ visited by residents of a given neighborhood $\gamma$, weighted by the number of trips $T$ between neighborhood $\gamma$ and place $\alpha$, leading to the following definition 

\begin{equation} \label{eq:2}
I_{\gamma} = \frac{\sum_{\alpha} (S_\alpha \times T_{\gamma\alpha})}{\sum_{\alpha} T_{\gamma\alpha}}.
\end{equation}

We then use Local Indicators of Spatial Autocorrelation (LISA) \cite{anselin1995local} to show the existence of larger zones of segregation and isolation, which allows us to understand how cities are structured. Spatial autocorrelation refers to the degree to which similar values cluster together in space \cite{getis2009spatial}, and this test lets us identify the clusters of contiguous zones where residents have limited exposure to diversity. LISA measures the correlation between an areal unit and its neighbors along some dimension; if values of segregation or isolation co-occur in space, we will see clustering. We use permutations, shuffling the values of our variables of interest and recomputing autocorrelation, to test the significance of each local I value, keeping only the values for which $p < 0.05$. The results can be used to not only to detect clusters of similar values, which we define as ``high-high" when the index cell and its neighbors are in the top third of segregation or isolation; we use ``low-low" for the opposite, as well as the remaining combinations of high and low mixes.  

\paragraph{Dimensionality reduction.} Our data span four years, from 2019 through 2022, which gives us the ability to assess the stability of these measures over time. To understand this, we both follow the raw trends and track relationships for a series of variables that predict diversity and exposure. We first build a time series of segregation and isolation for each neighborhood in our data, 48 observations (months) across 220,0000 neighborhoods (all Census block groups). Monitoring every time series presents a challenge, so we divide our data into 9 classes—combining 3 quantiles of segregation with 3 quantiles of isolation—at 2019 January $t_0$, and then assign classes according to the original breaks for each month $t_i$ thereafter, tracking the size of each class throughout our sample. We also compute average segregation and isolation for each metropolitan area and track those levels as well. Because our data represent a space-time cube, where each location in space has 48 values, we explore other forms of dimensionality reduction in the supplemental information. 

To understand the factors associated with these trends, we construct decision trees to predict segregation and isolation at each interval. We choose this modeling strategy because the patterns in our data involve symmetries and nonlinearities, with high and low extremes of a given variable giving similar predictions, and because variables appear to interact. A decision tree manages these challenges in a manner that preserves interpretability, as we can explore the tree to see how variables are stacked to generate predictions \cite{breiman2017classification}. The key tools that we use to interpret our data are the tree itself but also partial dependence plots, which show the best prediction given the values of two variables controlling for all others \cite{goldstein2015peeking}. 

\bibliography{mobility}

\section*{Acknowledgements}
The authors would like to thank the members of our research group who contributed comments throughout the process.

\section*{Author contributions statement}
A.R. collected data, wrote code, performed analysis and wrote the manuscript. N.O. and E.A. discussed results and edited the manuscript.  

\subsection*{Data and code availability}
We have made the processed data important to these findings, including measures of segregation and isolation, available at \href{https://asrenninger.github.io/diversity/}{this website}. For bulk downloads, please reach out the corresponding author; code will also be made available on request. 
\section*{Competing interests}
The authors declare no conflict of interest.

\end{document}


\flushbottom
\maketitle

\tableofcontents

\clearpage


\section{Complete decision tree}

\begin{figure}[h!]
\centering
\makebox[\textwidth][c]{\includegraphics[width=1.15\textwidth]{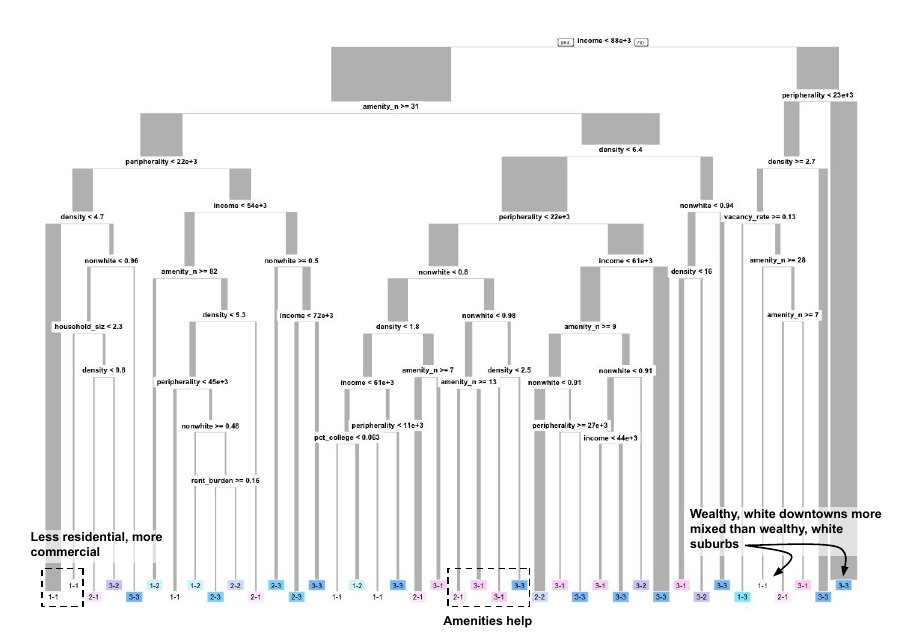}}
\caption{Decision tree with all predictors, showing that wealh, race, and location are dominant predictors, but also that amenities play an important role. There are segments of the population that avoid diversity, despite considerable variety in visitors nearby, but these groups are idiosyncratic, with a variety of features—that could be proxies for other facets of urban life—predicting them.}
\label{completetree}
\end{figure}

\clearpage

\section{Comparing income to nonwhite share}

\begin{figure}[h!]
\centering
\makebox[\textwidth][c]{\includegraphics[width=1.15\textwidth]{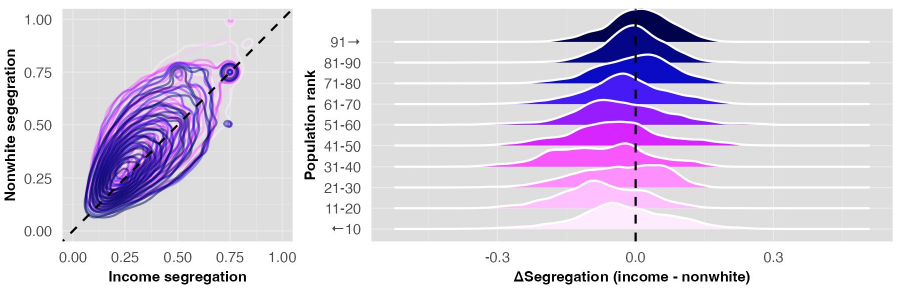}}
\caption{As a robustness check, we use estimates of the nonwhite population to check whether the relationships we see for segregation and isolation hold along dimensions other than income. In \textbf{A}, we show that place segregation using median income and place segregation using nonwhite population are correlated. In \textbf{B}, we subtract nonwhite segregation from income segregation for all observations and plot the distributions for different city sizes. For large cities, nonwhite segregation tends to be higher than income segregation, which strengthens our results. Comparison of segregation by estimated income and nonwhite population showing that although the measures are correlated, for larger cities by population, race segregation is systematically higher than class segregation.}
\label{racecheck}
\end{figure}

\clearpage

\section{NULL model}

\begin{figure}[h!]
\centering
\makebox[\textwidth][c]{\includegraphics[width=1.15\textwidth]{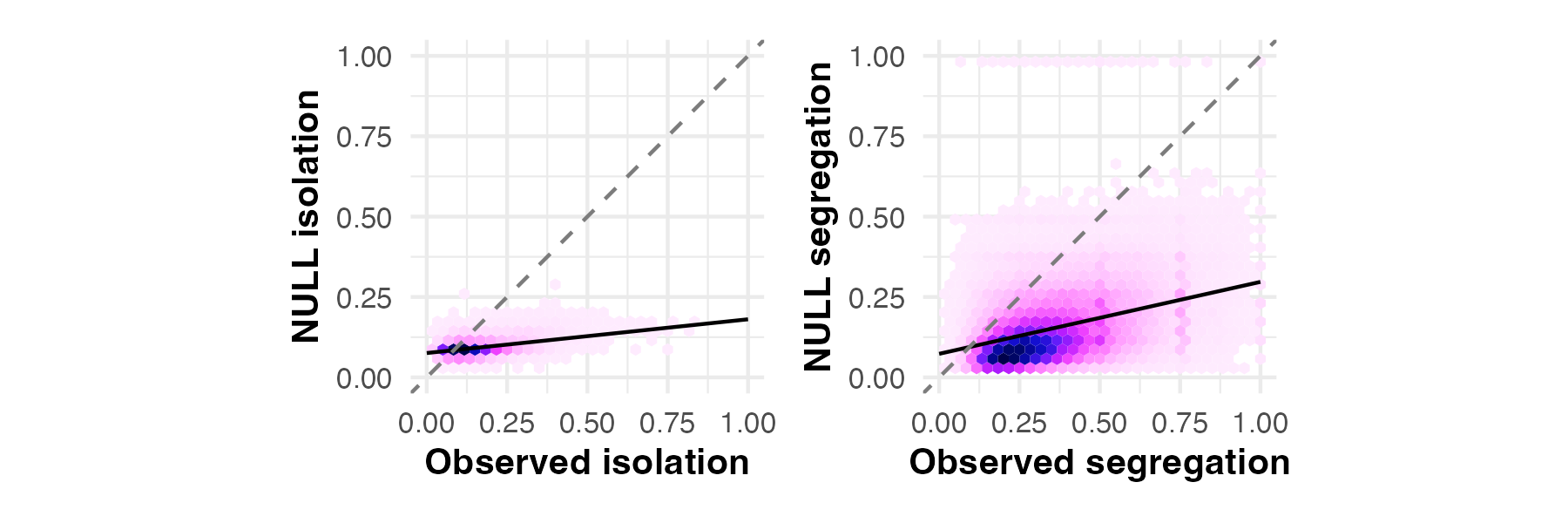}}
\caption{Mixing compared to a NULL model. We construct our measures of segregation and isolation using commuting data acquired from the Census Bureau's LEHD Origin-Destination Employment Statistics \cite{USCensusBureau_LEHD_LODES} We compute them for block groups and thus assume that all employees within one are exposed to each other. Segregation and isolation obtained from GPS data is systematically higher than when they are obtained from commute data under these assumptions, which suggests that people are sorting into amenities around where they work rather than mixing with the general population of workers. As the plot of isolation shows in particular, the assumption of general mixing is generous and there is less variation in these NULL results than in the GPS data, but it serves as a comparison to an unbiased counterfactual where individuals do not sort into amenities and simply mix and at home and at work.}
\label{lodes}
\end{figure}

\clearpage

\section{Analysis of ``churn"}

\begin{figure}[h!]
\centering
\makebox[\textwidth][c]{\includegraphics[width=1.15\textwidth]{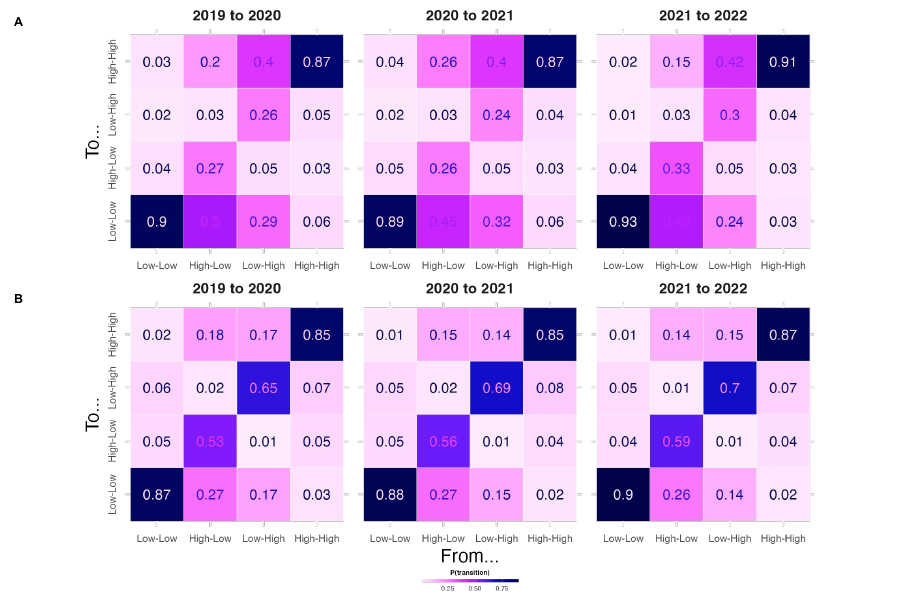}}
\caption{Transition matrices for segregated clusters (\textbf{A}) and isolated clusters (\textbf{B}) showing that the hotspots of high segregation and isolation are stable over time.}
\label{transition}
\end{figure}

\begin{figure}[h!]
\centering
\makebox[\textwidth][c]{\includegraphics[width=1.15\textwidth]{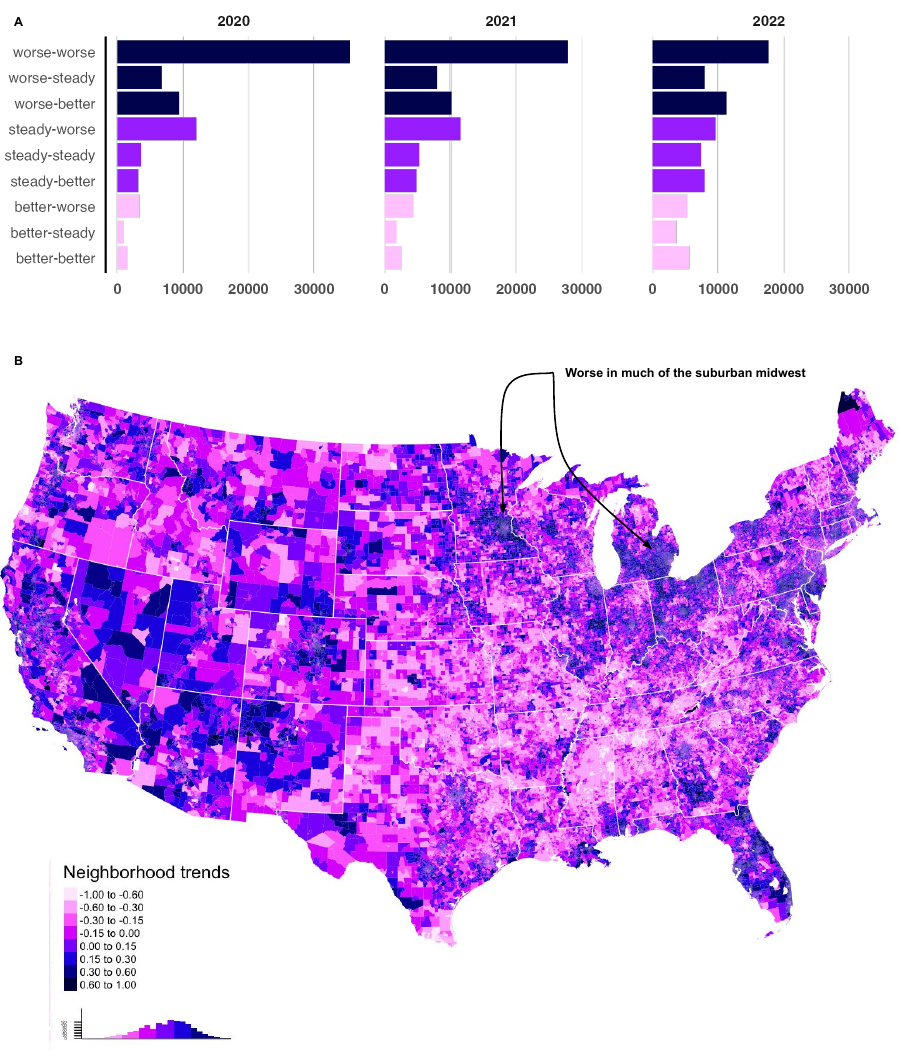}}
\caption{\textbf{A} Counts of different transitions between our 3x3 classes, along segregation and isolation (first-second) for the 20 largest metros: a plurality of neighborhoods have become more segregated and isolated. \textbf{B} Trends in isolation decomposed to neighborhoods show that certain parts of the country have experienced large increases; far fewer have seen decreases in isolation.}
\label{betterorworse}
\end{figure}

We reduce dimensionality by summarizing each neighborhood using the Mann-Kendall test for monotonicity \cite{mcleod2005kendall}. This test, which produces a statistic called Kendall's $\tau$, captures both the direction and strength of a trend: 1 indicates perfect monotonic increase and -1 perfect monotonic decrease, with monotonicity representing the degree to which a trend is consistent, each month building on the last, compared to up-and-down over time. Fig. \ref{betterorworse}A looks at the top 20 cities and shows that there is indeed a great deal of churn within top cities across the sample, with areas that became more isolated and more segregated in the plurality. This suggests that a rise in experienced segregation since the pandemic in 2020 is a big city phenomenon.Kendall's $\tau$, in Fig. \ref{betterorworse}B, shows that changes are also spatially clustered at the national level. In particular, the midwestern and northeastern parts of the country have areas that experienced steep rises in experience segregation.

\clearpage

\section{Tree stability}   

\begin{figure}[h!]
\centering
\makebox[\textwidth][c]{\includegraphics[width=1.15\textwidth]{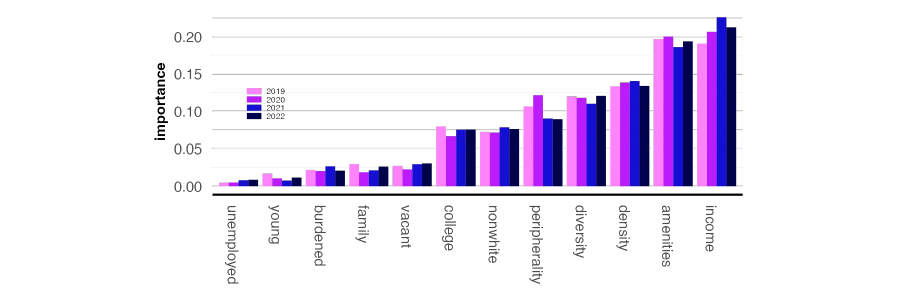}}
\caption{Feature importance over time showing remarkable stability, with the exception of a key variable: distance to the central business district. We also note that there is little change in rank importance.}
\label{importances}
\end{figure}

\clearpage

\bibliographystyle{naturemag}
\bibliography{mobility}